\documentclass[twocolumn,english,aps,showpacs,preprintnumbers,amsmath,amssymb,epsf]{revtex4}
\usepackage[latin1]{inputenc}
\usepackage{amsmath}
\usepackage{graphicx}
\usepackage{amssymb}

\makeatletter
\usepackage{amssymb}

\usepackage{amssymb}

\usepackage{amssymb}
\usepackage{graphicx}
\usepackage{dcolumn}
\usepackage{bm}

\usepackage{amssymb}

\usepackage{graphicx}

\begin{document}

\title{Observation of high-order quantum resonances in the kicked rotor}

\author{J.F. Kanem$^{1,2}$, S. Maneshi$^{1,2}$, M. Partlow$^{1,2}$, M.
Spanner$^{1,3}$ and A.M. Steinberg$^{1,2}$}

\affiliation{$^{1}$Centre for Quantum Information \& Quantum Control}

\affiliation{$^{2}$Institute for Optical Sciences \& Department of Physics, 
University of Toronto, 60 St. George Street, Toronto, Ontario, Canada M5S 1A7}

\affiliation{$^{3}$Chemical Physics Theory Group, Department of Chemistry, 
University of Toronto, 80 St. George Street, Toronto, Ontario, Canada M5S 3H6}

\date{April 7, 2006 }


\begin{abstract}
Quantum resonances in the kicked rotor are characterized by
a dramatically increased energy absorption rate, in stark contrast
to the momentum localization generally observed. These resonances
occur when the scaled Planck's constant $\tilde{\hbar}=\frac{r}{s}\cdot4\pi$,
for any integers $r$ and $s$. However only the $\tilde{\hbar}=r\cdot2\pi$
resonances are easily observable. We have observed high-order quantum
resonances ($s>2$) utilizing a sample of low temperature, non-condensed
atoms and a pulsed optical standing wave. Resonances are observed
for $\tilde{\hbar}=\frac{r}{16}\cdot4\pi$ for integers $r=2-6$.
Quantum numerical simulations suggest that our observation of high-order
resonances indicates a larger coherence length than expected from
an initially thermal atomic sample. 
\end{abstract}
\maketitle
A rotor subjected to a periodically pulsed sinusoidal potential ({}``kicked
rotor'') is one of the most widely studied paradigms of chaotic dynamics.
Ever since the qualitative differences between the classical kicked
rotor and the quantum kicked rotor (QKR) became evident \cite{casati},
the QKR has proven to be a rich system for studying quantum-classical
correspondence, decoherence, and quantum dynamics in general. To this
day the study of the standard QKR as well as alternative kicked rotor
Hamiltonians\cite{brumer-pre68-056202,brumer-pre70-016202} is an
actively pursued field. Much of the early work was done through theoretical
and numerical analysis, with one of the more important discoveries
being the realization that momentum localization in the QKR can be
thought of as a form of Anderson localization\cite{fishman-prl49-509}.
An experimental breakthrough in the field came when laser cooling
and optical trapping of atoms allowed the use of optical lattices
as a linear momentum analogue of the QKR. This led to the observation
of some of the theoretical predictions such as momentum 
localization\cite{raizen-prl75-4598}
as well as studies of decoherence\cite{raizen-pre61-7223} and interesting
results arising from modifications to the Hamiltonian of 
the QKR\cite{lignier-europhys69-327,kanem-mkr}.

Quantum resonances\cite{shepelyansky1,raizen-prl75-4598} are another
aspect of the QKR which have been of experimental interest recently:
for certain parameters, heating is greatly enhanced in contrast to
the momentum localization usually present in the quantum kicked rotor.
In the presence of gravity or other linear potentials one sees accelerator
modes\cite{oxford-pra62-013411}, similar to quantum resonances except
that there is an increase in average momentum as well as momentum
spread. Like other aspects of the quantum kicked rotor, quantum resonances
are useful for studying quantum-classical correspondence. Work has
gone into studying the effect in the presence of noise and the competition
with momentum localization and the resonances\cite{oxford-prl87-074102}.

Here we present experimental observation of quantum resonances utilizing
a sample of cold thermal rubidium atoms in an optical lattice. Specifically,
we report our observation of high-order quantum resonances. There
have been studies of high-order accelerator modes 
previously\cite{oxford-prl90-124102}
as well as a concurrent observation of high-order resonances in a
Bose-Einstein condensate\cite{phillips}. 
In all previous experiments with nondegenerate atoms,
the higher-order resonances were absent.

The Hamiltonian for the atom optics realization of the delta kicked
rotor is \begin{equation}
H=\frac{p^{2}}{2m}+\frac{U_{0}}{2}\left(1+\cos\left(2k_{L}x\right)\right)
\sum_{n}\delta\left(\frac{t}{T}-n\right)\end{equation}
 where $U_{0}$ is the strength of the sinusoidal kick, $2k_{L}$ is
the reciprocal lattice vector of the potential and $T$ 
is the period of the train
of delta kicks. It is convenient to express this as a scaled dimensionless
Hamiltonian:
\begin{equation}
\tilde{H}=\frac{\tilde{p}^{2}}{2}+\chi\left(1+\cos\left(\theta\right)\right)
\sum_{n}\delta\left(\tilde{t}-n\right)\end{equation}
 where $\tilde{p}=2Tk_{L}p/m$, $\theta=2k_{L}x$, $\tilde{t}=t/T$,
$\chi=2U_{0}k_{L}^{2}T^{2}/m$, and $\tilde{H}=H4T^{2}k_{L}^{2}/m$.

The scaled quantum Schr\"{o}dinger's equation is \begin{equation}
\imath\tilde{\hbar}\frac{\partial}{\partial\tilde{t}}\psi=-\frac{\tilde
{\hbar}^{2}}{2}\frac{\partial}{\partial^{2}\theta}\psi+\chi\left(1+\cos\theta
\right)\sum_{n}\delta\left(\tilde{t}-n\right)\psi\label{sse}\end{equation}
 where $\tilde{\hbar}=4Tk_{L}^{2}\hbar/m$ is the scaled Planck's
constant. This effective Planck's constant is a measure of the magnitude
of the quantized momentum transfer due to the lattice ($2\hbar k_{L}$),
relative to the momentum required to move one lattice spacing in one
kick period, $T$. Its value determines how quantum-mechanically the
system behaves and whether or not quantum resonances are observed.

Quantum resonances occur when $\tilde{\hbar}=\frac{r}{s}\cdot4\pi$,
where $r$ and $s$ are integers. Note that for a given experimental
setup $\tilde{\hbar}$ is only sensitive to the kick period, $T$, which
can be controlled with great precision.
These resonances can be thought of as a rephasing of the momentum
states coupled by the lattice potential, whose momenta differ by a
multiple of $2\hbar k_{L}$. Indeed, in the delta kick limit, the
condition above can be found by setting the phase between two states
accumulated between successive delta kicks to some integer multiple
of $2\pi$: $\Delta\phi_{a}-\Delta\phi_{b}=\left(a^{2}-b^{2}\right)2\hbar k_
{L}^{2}T/m=\left(a^{2}-b^{2}\right)\frac{\tilde{\hbar}}{2}=q\cdot2\pi$,
or $\tilde{\hbar}=\frac{q}{\left(a^{2}-b^{2}\right)}4\cdot\pi=\frac{r}{s}
\cdot4\pi$,
recognizing that for any integers $a$ and $b$, $\left(a^{2}-b^{2}\right)$
is also an integer. When this is satisfied for $s=1$, all coupled
momentum states rephase. These first order quantum resonances are
related to the revivals of the wavepackets; if the time between kicks
is a multiple of the revival time, all the kicks add coherently. This
leads to a linear growth in the width of the momentum distribution
and quadratic energy growth with kick number, $n$\cite{quadratic}.
This is in contrast to the chaotic situation which arises when the
position of the particle for successive kicks is essentially uncorrelated
and the energy growth is linear. For the case of $s>1$, the resulting
high-order resonances are a manifestation of the fractional revivals,
in which some but not all of the eigenstates rephase, and the wavepacket
recoalesces, split into $s$ identical copies.

In the experimental realization the delta kick train is replaced with
a train of square pulses of finite width $t_p$:
$\Delta_{n}\left(\tilde{t}\right)=\Theta\left(\tilde{t}-n\right)-\Theta\left
(\tilde{t}-\frac{t_{p}}{T}-n\right)$,
where $\Theta\left(\tilde{t}\right)$ is the Heaviside step function.
Equation (\ref{sse}) then becomes
\begin{equation}
\imath\tilde{\hbar}\frac{\partial}{\partial\tilde{t}}\psi=-\frac{\tilde
{\hbar}^{2}}{2}\frac{\partial}{\partial^{2}\theta}\psi+\kappa\frac{T}{t_{p}}
\left(1+\cos\theta\right)\sum_{n}\Delta_{n}\left(\tilde{t}\right)\psi\end
{equation}
 where $\kappa=2Tk_{L}^{2}V_{0}t_{p}/m$ is the stochasticity parameter
and $V_{0}$ is the energy depth of the lattice potential. In the
classical kicked rotor this stochasticity parameter completely defines
how chaotic the behavior of the system is.

There are also classical resonances which result in increased energy
absorption of the system. A key difference is the condition required;
$\kappa=\sqrt{\left(n\cdot2\pi\right)^{2}+16}$ for integer $n>0$\cite
{lieberman,ishizaki},
showing a dependence not solely on the kick period, but also on the
strength of the kicks, indicating a Newtonian effect completely independent
of $\tilde{\hbar}$.

Our optical lattice is formed in the vertical direction by two laser
beams of wavelength $\lambda=780 \, {\rm nm}$ intersecting at an angle of $
\gamma=49.0^{\circ}\pm0.2^{\circ}$
resulting in a laser intensity interference pattern of the form $I\left(x
\right)=I_{0}\cos^{2}\left(\frac{2\pi}{\lambda}x\sin\left(\gamma/2\right)
\right)$.
The lattice is detuned by $\Delta\simeq2\pi\cdot20 \, {\rm GHz}$ from the 
$F=3\Rightarrow F^{\prime}=4$
D2 trapping line of $^{85}Rb$, far enough to be treated as a conservative
potential. The spatial lattice period is $l=\frac{\lambda}{2\sin\left
(\gamma/2\right)}=0.940 \, \mu {\rm m}\pm0.004 \, \mu {\rm m}$.
The resulting light shift on the atoms produces a potential $V\left(x\right)
=I\left(x\right)\frac{\hbar\Gamma^{2}}{4\Delta I_{s}}$
where $\Delta$ is the detuning of the laser from the atomic resonance
and $\Gamma$ and $I_{s}$ are the natural linewidth and saturation
intensity, respectively , of the atom. In accordance with the kicked
rotor model this potential can be expressed as $V\left(\theta\right)=V_{0}
\left(1+\cos\theta\right)$
where $V_{0}=\frac{I_{0}}{I_{s}}\frac{\hbar\Gamma^{2}}{8\Delta}$,
$\theta=2k_{L}x$ and $k_{L}=\frac{2\pi}{\lambda}\sin\left(\gamma/2\right)$.

We prepare for our experiment by loading a shallow optical lattice
from a magneto-optical trap (MOT) of rubidium. While the MOT is still
present the lattice laser beams are turned on so that subsequent molasses
cooling is done in the presence of the lattice, leading to a higher
loading efficiency\cite{weiss-prl82-2263}. This initial lattice is
tailored to have a depth of $\sim19E_{r}=k_{B}\cdot605  \, {\rm nK}$ where 
$E_{r}=\frac{\hbar^{2}k_{L}^{2}}{2m}
\approx k_{B}\cdot32  \, {\rm nK}$
is the effective recoil energy. For our lattice geometry this depth
supports two bound states. Because of the vertical orientation of
the lattice any unbound atoms will fall out of the interaction region
due to gravity. After $\gtrsim10  \, {\rm ms}$ only atoms in the two bound 
states
remain, at kinetic temperatures of $\left(\Delta p\right)^{2}/k_{B}\sim120  
\, {\rm nK}$
and $\sim360  \, {\rm nK}$ for the ground and first excited states 
respectively,
making for a very cold sample of atoms ($\sim240  \, {\rm nK}$). Time-of-
flight
measurements are limited by the time it takes for the atomic sample
to fall out of the observation region. As a result, this $\sim240  \, {\rm 
nK}$
is the upper bound which we can place experimentally on the initial
atom cloud temperature. This low temperature is to be contrasted with
the experiments of most previous work on the subject of quantum resonances
which have a molasses-cooled atomic cloud of temperature on the order
of several microKelvins. Since the coherence length in a molasses,
from which our lattice is loaded, is smaller than an optical wavelength,
we do not expect any coherence between lattice wells despite this
low temperature. Within each well there is an incoherent mixture of
the two bound states. In other words the bands of the lattice are
expected to be completely and incoherently filled.

The presence of gravity also adds a tilt to the potential of $3.0E_{r}$
per spatial lattice period. This tilt would completely alter the dynamics
of the kicked rotor in a manner which is 
interesting\cite{oxford-pra62-013411}
but quite different from the theory outlined above. To overcome this
we accelerate the lattice downward at $g$, putting the whole experiment
in free fall, thus getting rid of the potential tilt. This is done
by frequency-shifting each of the lattice beams by acousto-optic modulators
(AOMs). The frequency of the radiofrequency (RF) power driving each
AOM is independently controlled by a programmable function generator.
By linearly ramping the relative frequency difference between each
RF signal, we can accelerate the lattice. The relative frequency difference
actually undergoes discrete frequency jumps of $105  \, {\rm Hz}$ 
every $10 \, \mu {\rm s}$.
Trial reductions of the granularity by reducing the time between frequency
jumps to $5 \, \mu {\rm s}$ and then $1 \, \mu {\rm s}$ had 
no effect on the results.
The uncertainty in the lattice angle between lattice beams translates
to an uncertainty in the acceleration being applied to the lattice
of $0.07  \, {\rm m/s}^{2}$.

\begin{figure}
\includegraphics[%
  width=8.5cm,
  keepaspectratio]{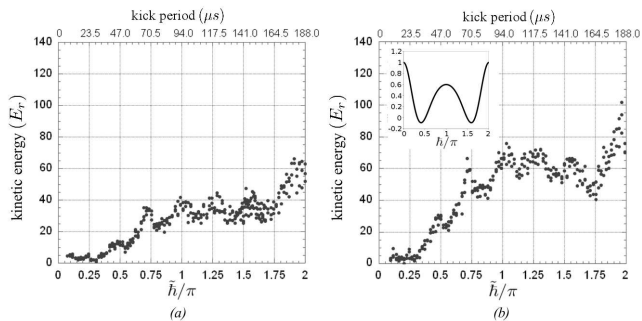}

\caption{Experimental graphs showing energy in recoil energies, $E_{r}$,
versus $\tilde{\hbar}/\pi$, and period, $T$, both with a kick number
of $16$ $\left(a\right)$ Kick strength of $k=1.23$. High order
resonances are seen for $\tilde{\hbar}=\frac{r}{16}\cdot4\pi$ for
$r=2-6$. $\left(b\right)$ Same as $\left(a\right)$ but with a kick
strength of $k=1.64$. Background trend is from a general kick strength
dependent quantum diffusion rate and is modeled by equation \ref{diff}
which is shown in figure in the inset for $k=1.6$.}
\end{figure}

The temporal modulation of the lattice potential for the pulse train
is done by modulating the laser intensity, also controlled with the
AOMs. The RF signals to the AOMs are sent through RF switches, which
are also controlled with a programmable function generator, allowing
for a rise/fall time of the lattice intensity of $\sim200  \, {\rm ns}$. The
width of the kicks, $t_{p}$, ranges from $5$ to $10 \, \mu {\rm s}$ and is
used as a control for the kick strength, 
$k\equiv\kappa/\tilde{\hbar}=\frac{V_{0}t_{p}}{2\hbar}$.
Simulations show a very small difference between using true delta
pulses and the relatively wide pulses we use for the actual experiment,
while the experiment itself has confirmed that there is little effect
of varying the pulse width but keeping $k$ constant.

After a train of kicks is applied to the atomic ensemble we perform
a time-of-flight measurement. First the ensemble undergoes $32  \, {\rm ms}$
of free expansion, and then an image is recorded by flashing resonant
light onto the sample and collecting the scattered light on a CCD
camera. From this image the thermal energy of the sample is extracted.

Figure 1a shows a typical graph of the average energy of the atoms
after $16$ kicks versus kick period and effective Planck's constant.
The depth of the kicking potential in this case was $V_{0}=98E_{r}$
with a pulse width of $t_{p}=6 \, \mu {\rm s}$ resulting in a kick strength
of $k=\frac{V_{0}t_{p}}{2\hbar}=1.23$. Figure 1b is a similar graph
but with $t_{p}=8 \, \mu {\rm s}$ resulting in a kick strength of $k=1.64$.

The general background trend is a kick strength effect independent
of the high-order resonances and can be modeled by the quantum diffusion
parameter \begin{equation}
D\left(k,\tilde{\hbar}\right)=\frac{k^{2}\tilde{\hbar}^{2}}{2}\left[\frac{1}
{2}-J_{2}\left(d\right)-J_{1}^{2}\left(d\right)+J_{2}^{2}\left(d\right)+J_{3}
^{2}\left(d\right)\right]\label{diff}\end{equation}
 where $J_{n}$ are $nth$ order Bessel functions and $d\equiv2k\sin\left
(\tilde{\hbar}/2\right)$\cite{lieberman,shepelyansky2}.
The energy increase due to this diffusion is given by $\frac{m}{4k_{L}^{2}T^
{2}}D\left(k,\tilde{\hbar}\right)$.
The result is shown in the inset of figure 2b. This equation does
not model the decreasing stochasticity ($\kappa=k\tilde{\hbar}$)
as $\tilde{\hbar}\rightarrow0$, explaining its failure to reproduce
the vanishing heating rate seen in the experiment.

The experiment was performed for several different parameters in order
to verify that the position of the resonances is solely a function
of the effective Planck's constant. The kick strength was varied by
adjusting the temporal width of the pulses from $6$ to $14 \, \mu {\rm s}$,
leading to kick strengths from $k=1.23$ to $k=2.86$. We see that
an increase by more than a factor of two, which would displace classical
resonances by the same factor, leaves the high order resonance structure
in place to within our experimental error of about $\pm2 \, \mu {\rm s}$. 
These
resonances lie at an average period of $T=45.2\pm0.9$, $69.0\pm1.0$,
$96.0\pm1.4$, $120.0\pm1.4$, and $147.5\pm1.0 \, \mu {\rm s}$. These 
correspond
to values of $\tilde{\hbar}/\pi=0.47\pm0.01$, $0.72\pm0.01$, $1$,
$1.25\pm0.02$, $1.54\pm0.02$. (The position of the central resonance is
defined to be precisely $1$ because it itself is
our most accurate calibration of $\tilde{\hbar}$.)  We identify these as the 
high order
resonances occurring for $\tilde{\hbar}=\frac{r}{16}\cdot4\pi$, for
$r=2-6$.

We also examined the characteristics of the resonances as a function
of kick number. The primary trend noted is the clarity of the resonances.
Figure 2 shows peak height in energy (over the baseline) of the $\tilde
{\hbar}=\frac{3}{4}\pi$
resonance for different kick numbers. In other words, this is the
kinetic energy at the resonance minus the energy minimum between resonances.
We believe this reflects the narrowing of the resonances with increased
kick number.

\begin{figure}
\includegraphics[%
  width=6cm,
  keepaspectratio]{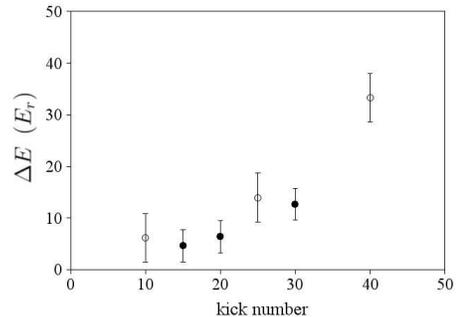}

\caption{Magnitude of the $\tilde{\hbar}=\frac{3}{4}\pi$ resonance as a 
function
of kick number. Solid and open circles indicate different sets of
experimental data.}
\end{figure}

We have studied these quantum resonances numerically while modeling
the experiment with realistic parameters. Figure 3 shows the results
of quantum simulations graphing energy absorption in recoil energies
after $16$ kicks versus kick period (and $\tilde{\hbar}/\pi$) for
$k=1.6$. There is a strong sharp resonance at $\tilde{\hbar}=2\pi$
as well as a higher order one at $\tilde{\hbar}=\pi$, and less well
resolved high-order resonances at $\tilde{\hbar}\sim\frac{3}{4}\pi$,
among others at $\tilde{\hbar}\sim\frac{r}{s}\cdot4\pi$ of order
$s>16$. The initial state of 3a is an uncertainty-limited Gaussian
momentum distribution, $\psi\left(x\right)=\left(\frac{1}{2\pi x_{0}^{2}}
\right)^{1/4}\exp\left[-\frac{x^{2}}{4x_{0}^{2}}\right]$
with velocity width of $v_{0}=\frac{\hbar}{2mx_{0}}=3.43  \, {\rm mm/s}$. 
This
is the velocity distribution of the ground band of a $19E_{r}$ deep
lattice, a state we can readily prepare\cite{kanem-phasespace}. This
gives a spatial width of $x_{0}=0.109 \, \mu {\rm m}$ which localizes each
wavepacket entirely within one well. There is no defined phase between
wavefunctions in adjacent wells.

\begin{figure}
\includegraphics[%
  width=8.5cm,
  keepaspectratio]{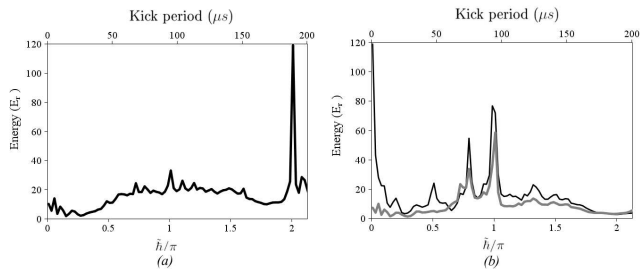}

\caption{Numerical simulations showing energy in recoil energies, $E_{r}$,
versus $\tilde{\hbar}/\pi$, and period, $T$, both with a kick number
of $16$ and $k=1.6$. $\left(a\right)$: The $rms$ spatial distribution
of the initial wavefunction is 
$x_{0}=0.109 \, \mu {\rm m}$ ($v_{0}=3.43  \, {\rm mm/s}$).
$\left(b\right)$: Initial distributions with coherence throughout
the sample. The grey line has an initial distribution of a sum of
in-phase Gaussians, each separated by the lattice spatial period,
$l=0.94 \, \mu {\rm m}$, while that of the black line has a wide 
initial $rms$
spatial width of $x_{0}=1.09 \, \mu {\rm m}$. \protect \\
}
\end{figure}
We find that the initial conditions of the atomic cloud, namely its
momentum distribution and coherence properties, have a large effect
on the relative strengths of the various high-order resonances. Figure
3b shows the results of two numerical simulations, each having an
initial distribution with a coherence length spanning several lattice
spatial periods. The initial distribution of the grey line is a sum
of Gaussians, weighted by an overall Gaussian envelope. Each individual
Gaussian has a spatial distribution width of $x_{0}=0.109 \, \mu {\rm m}$ as
in 3a. This makes for a constant phase from well to well throughout
the atomic sample; $\psi_{0}\left(x\right)=N\exp\left[-\frac{x^{2}}{4x_{r}^
{2}}\right]\sum_{n}\exp\left[-\frac{\left(x-n\cdot l\right)^{2}}{4x_{0}^{2}}
\right]$,
where $N$ is a normalization constant and $x_{r}=1.09 \, \mu {\rm m}$ is the
$rms$ radius of the overall Gaussian that each individual {}``cloud''
is weighted by. The black line is a single Gaussian with an initial
spatial $rms$ width of $x_{0}=1.09 \, \mu {\rm m}$ making for a velocity 
distribution
of $rms$ width $v_{0}=0.343  \, {\rm mm/s}$. The result of these 
distributions
show clearly observable resonances. They are visible at 
$\tilde{\hbar}=\frac{r}{16}\cdot4\pi$
for $r=2,3,4$ with another one at $\tilde{\hbar}=\frac{1}{5}\cdot4\pi$.
The other obvious difference between figures 3b and 3a is the absence
of the resonance at $\tilde{\hbar}=2\pi$. It is worth noting that
in an angular momentum kicked rotor system, as opposed to the optical
lattice implementation used here, this resonance at $\tilde{\hbar}=2\pi$
is never present. This is one of several subtle differences owing
to the fact that in the kicked rotor the wavefunction must be cyclic;
$\psi\left(\theta\right)=\psi\left(\theta+2\pi\right)$. This is only
true in the linear momentum analogue of the optical lattice if there
is coherence throughout the atomic sample. The initial distribution
of figure 3b has this long range coherence put in by hand for the
simulation shown in grey, while the extremely low initial momentum
spread of the simulation shown in black makes for a long spatial coherence. 

It is also worth noting that many small higher-order resonances exist
and appear in figure 3a none of them dominating the others. But when
the initial distribution has coherence throughout several wells the
$\tilde{\hbar}=\frac{r}{16}\cdot4\pi$ resonances seen in figures
3b grow several times stronger than the others, in qualitative agreement
with the experimental observations shown in figures 1a,b.
The fact that the numerical simulations do not show clear high-order
resonances unless there is long-range coherence suggests that our
initial atomic state actually has some phase-coherence between adjacent
wells; we are currently investigating this possibility experimentally
and theoretically.

In summary we have observed clear high-order resonances
in an optical-lattice implementation of the kicked rotor using thermal
atoms. Our technique of selecting an extremely cold sample of atoms
may explain the fact that we observe these resonances, not previously
seen in thermal clouds. We have verified, in accordance with theory,
that only the value of the effective Planck's constant determines
the positions of the resonances. In addition our results suggest
that there may be long range coherence across the optical lattice,
which was loaded from an ensemble of thermal atoms. Future studies
will examine the role of inter- and intra-well coherence in the kicked
rotor as well as the various mechanisms for such coherence to be developed.
Our findings also suggest that the QKR may be of interest as a probe
of such coherence.

We would like to thank Paul Brumer and Jiangbin Gong for inspiration
and useful discussions. This work was funded by NSERC and PRO.


\begin{references}
\bibitem{quadratic} A quadratic energy growth with kick number is actually 
only seen in the angular momentum verison of the kicked rotor. The linear 
momentum analogue utilized here shows a linear energy growth until momentum 
localization sets in.
\end{references}
\end{document}